%% file: main.tex
\documentclass[sigconf, natbib=true, screen]{acmart}
\usepackage[linesnumbered,lined]{algorithm2e}
\usepackage{booktabs}
\usepackage{bookmark,xspace}
\usepackage{multirow}
\usepackage{paralist}
\usepackage{graphicx}
\usepackage{subfigure}
\usepackage{amsmath}
\usepackage{bbding}
\usepackage{enumitem}
\usepackage{color}
\usepackage{makecell}

\AtBeginDocument{%
  \providecommand\BibTeX{{%
    \normalfont B\kern-0.5em{\scshape i\kern-0.25em b}\kern-0.8em\TeX}}}

\newcommand{\name}{E4SRec\xspace}

\copyrightyear{2023}
\acmYear{2023}
\setcopyright{acmcopyright}
\acmConference[WWW '24]{Proceedings of the ACM Web Conference 2024}{May 13--17, 2024}{Singapore, Singapore}
\acmBooktitle{Proceedings of the ACM Web Conference 2024 (WWW '24), May 13--17, 2024, Singapore, Singapore}
\acmPrice{15.00}
\acmISBN{978-1-4503-XXXX-X/18/06}
\acmDOI{10.1145/1122445.1122456}

\begin{document}

\title{E4SRec: An Elegant Effective Efficient Extensible Solution of Large Language Models for Sequential Recommendation}

\author{Xinhang Li}
\affiliation{%
 \institution{Tsinghua Univerisity}
 \city{Beijing}
 \country{China}
}
\email{xh-li20@mails.tsinghua.edu.cn}

\author{Chong Chen}
\affiliation{%
 \institution{Huawei Cloud BU}
 \city{Beijing}
 \country{China}
}
\email{chenchong55@huawei.com}

\author{Xiangyu Zhao}
\affiliation{%
 \institution{City Univerisity of Hong Kong}
 \country{Hong Kong}
}
\email{xianzhao@cityu.edu.hk}

\author{Yong Zhang}
\affiliation{%
 \institution{Tsinghua Univerisity}
 \city{Beijing}
 \country{China}
}
\email{zhangyong05@tsinghua.edu.cn}

\author{Chunxiao Xing}
\affiliation{%
 \institution{Tsinghua Univerisity}
 \city{Beijing}
 \country{China}
}
\email{xingcx@tsinghua.edu.cn}

\renewcommand{\shortauthors}{Xinhang Li et al.}

\input{src/0-abs}
\maketitle
\input{src/1-intro}
\input{src/3-method}

\input{src/4-exper}
\input{src/2-relatedworks}
\input{src/5-conclusion}

\bibliographystyle{ACM-Reference-Format}
\bibliography{custom}

\input{src/6-appendix}

\end{document}

%% file: src/0-abs.tex
\begin{abstract}

The recent advancements in Large Language Models (LLMs) have sparked interest in harnessing their potential within recommender systems. Since LLMs are designed for natural language tasks, existing recommendation approaches have predominantly transformed recommendation tasks into open-domain natural language generation tasks. However, this approach necessitates items to possess rich semantic information, often generates out-of-range results, and suffers from notably low efficiency and limited extensibility.
Furthermore, practical ID-based recommendation strategies, reliant on a huge number of unique identities (IDs) to represent users and items, have gained prominence in real-world recommender systems due to their effectiveness and efficiency. Nevertheless, the incapacity of LLMs to model IDs presents a formidable challenge when seeking to leverage LLMs for personalized recommendations.
In this paper, we introduce an \textbf{E}legant \textbf{E}ffective \textbf{E}fficient \textbf{E}xtensible solution for large language models for \textbf{S}equential \textbf{Rec}ommendation (\textbf{\name}), which seamlessly integrates LLMs with traditional recommender systems that exclusively utilize IDs to represent items. 
Specifically, \name takes ID sequences as inputs, ensuring that the generated outputs fall within the candidate lists. Furthermore, \name possesses the capability to generate the entire ranking list in a single forward process, and demands only a minimal set of pluggable parameters, which are trained for each dataset while keeping the entire LLM frozen. We substantiate the effectiveness, efficiency, and extensibility of our proposed \name through comprehensive experiments conducted on four widely-used real-world datasets. The implementation code is accessible at https://github.com/HestiaSky/E4SRec/.




\end{abstract}

\ccsdesc[500]{Information systems~Recommender systems}
\ccsdesc[500]{Information systems~Language models}
\ccsdesc[500]{Information systems~Personalization}

\keywords{Large Language Model, Sequential Recommendation, Item ID and Indexing}

%% file: src/1-intro.tex
\section{Introduction}

\begin{figure}[t]
    \centering
    \includegraphics[width=\linewidth]{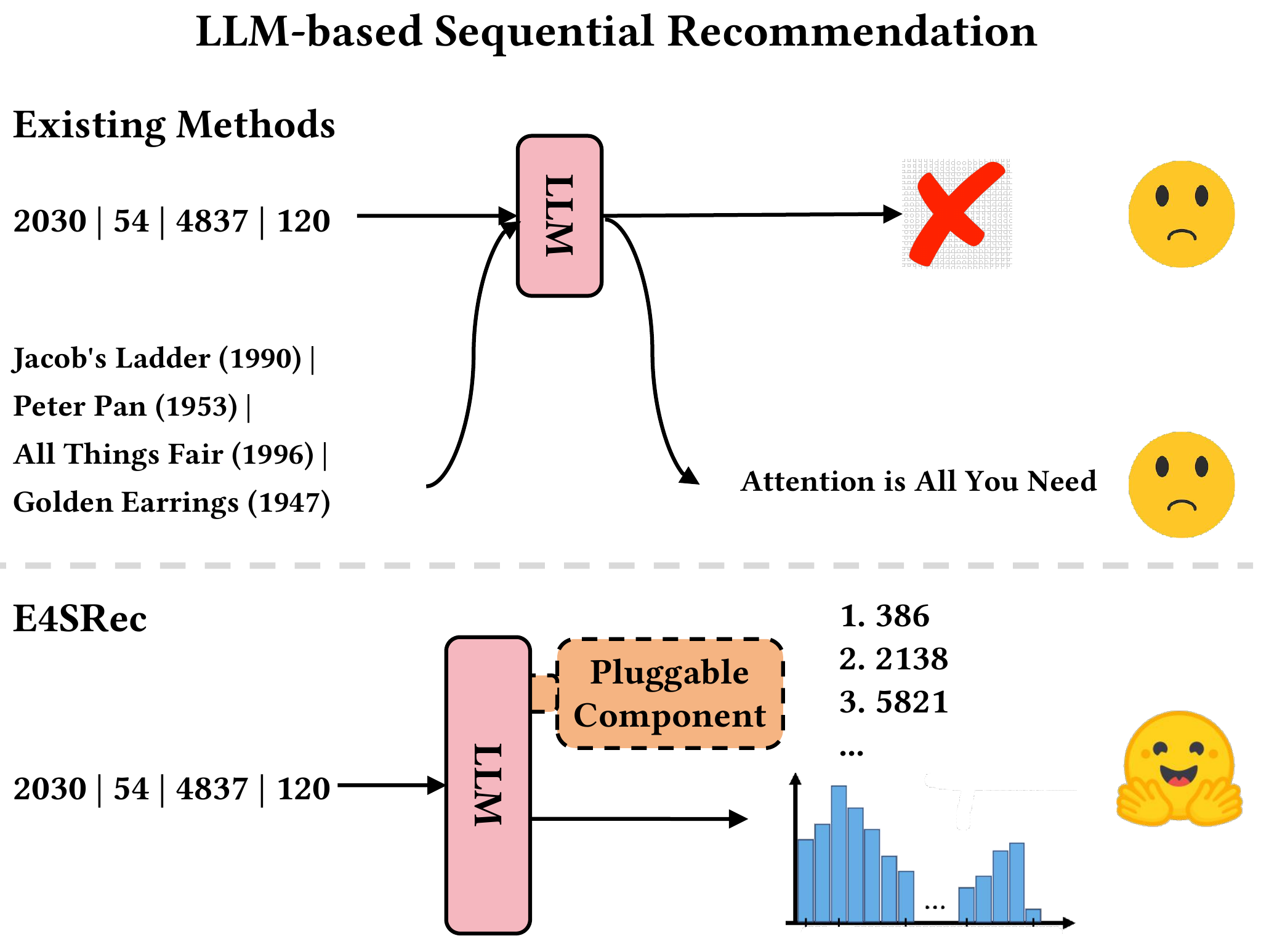}
    \caption{Illustration of LLM-based sequential recommendation. The upper part denotes the existing methods that fail to model the IDs and undesirably generate out-of-range results while the lower part denotes \name that can effectively and efficiently handle the IDs.}
    \label{fig:illustration}
\end{figure}

Recommender systems, in existence for decades, are instrumental in mitigating information overload and enhancing user experience on the Web~\cite{DBLP:conf/cscw/ResnickISBR94,DBLP:journals/cacm/KonstanMMHGR97,DBLP:journals/tkde/AdomaviciusT05}.
These systems discern user preferences to offer tailored recommendations on content or items~\cite{DBLP:conf/www/SarwarKKR01,DBLP:journals/internet/LindenSY03,DBLP:journals/tois/DeshpandeK04}.
Nowadays, the rise of Large Language Models (LLMs)~\cite{DBLP:conf/nips/KojimaGRMI22} is revolutionizing our familiar landscapes~\cite{vanDis2023ChatGPTFP}, which excel in assimilating real-world knowledge from the Web and achieving proficient natural language generation.
Recently, there has been a significant upsurge in research endeavors focused on leveraging LLMs for recommendation tasks and this trend is progressively becoming more inevitable~\cite{DBLP:journals/corr/abs-2305-19860}.
While many non-tuning approaches~\cite{DBLP:journals/corr/abs-2303-14524,DBLP:journals/corr/abs-2304-10149,DBLP:journals/corr/abs-2305-08845}, including prompting and in-context learning~\cite{DBLP:conf/nips/BrownMRSKDNSSAA20}, strive to leverage the zero/few-shot learning ability, tuning approaches usually outperform them as they are fine-tuned for specific tasks using dedicated data.
However, bridging the substantial gap between natural language generation tasks and recommendation tasks remains a formidable challenge.

To tackle above challenge, existing approaches~\cite{DBLP:conf/recsys/Geng0FGZ22,DBLP:conf/recsys/BaoZZWF023,DBLP:journals/corr/abs-2305-07001,DBLP:journals/corr/abs-2307-00457} predominantly convert the recommendation task into a natural language generation task to align it with the inherent capabilities of LLMs.
This involves the direct generation of item names or ratings based on appropriate prompts.
However, such a solution has several limitations as depicted in Figure~\ref{fig:illustration}.
First, these methods aim to harness the inherent knowledge of LLMs for recommendation through fine-tuning, essentially crafting an external knowledge-augmented content-based recommendation~\cite{DBLP:journals/corr/abs-2305-19860}.
Hence, they demand rich semantic information through the prompt.
When the semantic information is insufficient or vague due to a huge number of homogeneous items, which is very common in recommendation, these methods are not able to yield satisfying performance.
Different from content-based filtering methods, collaborative filtering methods utilize a huge number of unique identities (IDs) to represent users and items and have dominated the recommender system area for years.
Such IDs merely function as indices for users or items without encapsulating any semantic information.
Consequently, existing LLM-based recommendation methods struggle to manage the IDs to conduct ID-based recommendation and more critically, fail to leverage collaborative data vital for recommendations.
Second, the problem definition of the existing approaches that is generate the results from the whole vocabulary will often lead to out-of-range results~\cite{DBLP:journals/corr/abs-2307-00457}.
Such erratic generation not only diverges from the expectations of recommender systems but also negatively impacts user experience.
Last but not least, the existing methods are only able to generate one recommendation result each time for the characteristics of LLMs.
They mainly focus on the ranking task by attaching all the candidates in the prompt~\cite{DBLP:journals/corr/abs-2305-07001}, and cannot tackle the matching task that requires matching scores of all the candidates.
Nevertheless, the efficiency is still unacceptable for the recommender systems that require low latency and high concurrency.
Hence, the trajectory of current LLM-based recommendation techniques fails to meet the needs of contemporary recommender systems and lacks practicality.

To overcome the aforementioned limitations, we propose an \textbf{E}legant \textbf{E}ffective \textbf{E}fficient \textbf{E}xtensible solution of large language models for \textbf{S}equential \textbf{Rec}ommendation (\textbf{\name}) by incorporating LLMs and traditional recommendation models with only IDs to represent items.
Specifically, our proposed \name accepts only ID sequences as inputs and ensures controllable generation with high efficiency by making predictions on and only on all the candidates in each forward process.
Our proposed \name solution encompasses four key phases: sequential recommendation model pretraining, LLM instruction tuning, \name model training and \name model deployment.
For each given sequential recommendation dataset, we first pretrain a traditional sequential recommendation model and then extract the item ID embeddings to prepare for the ID injection of LLM.
An instruction tuning process of the LLM is also carried out to stimulate its capability to follow instructions and this tuned LLM is shared for all the task-specific models.
Then, in the training stage of \name, we wrap the sequences of item IDs into prompts by a linear projection of the item ID embeddings for ID injection.
We freeze all the parameters of the LLM and only train an additional minimal set of parameters for adaption on the specific dataset.
The recommendation results are made by computing the joint probability distribution between the output of LLM and all the candidate items via an item linear projection.
Finally, once being trained, \name can be deployed for practical application in a lightweight manner necessitating merely four pluggable components: the item ID embeddings, the input linear projection, the adapter and the item linear projection.

The contribution of this paper can be summarized as follows:
\begin{itemize}[leftmargin=*]
    \item We pioneer an innovative and effective strategy to address the unique challenges of integrating IDs in applying LLMs for recommendation tasks.
    \item We address the prevailing issues of out-of-range outputs and generation efficiency, achieving controllable and efficient generative recommendation.
    \item We propose an \textbf{E}legant \textbf{E}ffective \textbf{E}fficient \textbf{E}xtensible solution of large language models for \textbf{S}equential \textbf{Rec}ommendation (\textbf{\name}), which is able to build an industrial-level recommender system from scratch.
    \item Comprehensive experiments across four prominent real-world sequential recommendation datasets demonstrate the superiority and effectiveness of our proposed \name model with in-depth analyses underscoring its efficiency and extensibility in real-world applications.
\end{itemize}

%% file: src/3-method.tex
\begin{figure*}
    \centering
    \includegraphics[width=\linewidth]{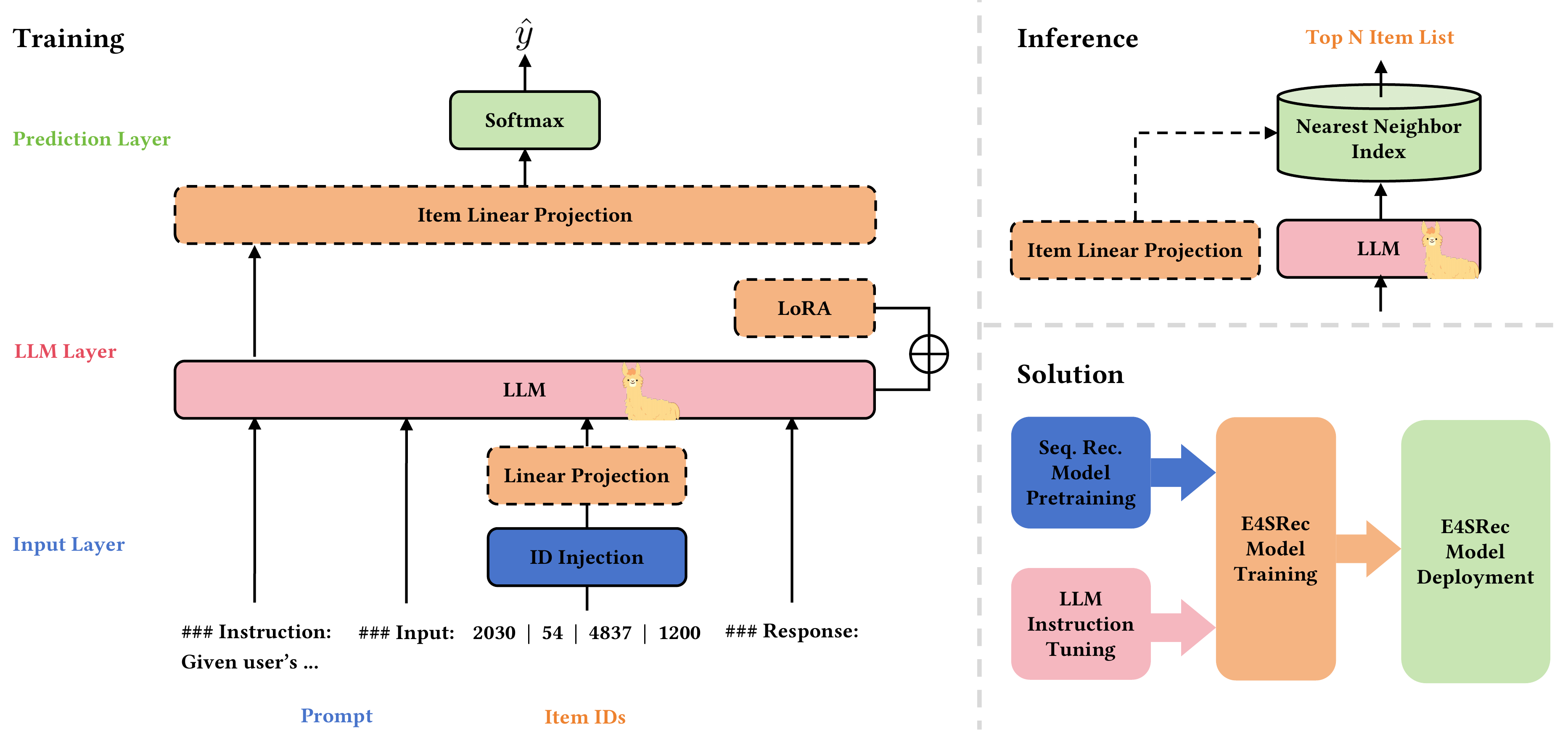}
    \caption{Architecture of \name solution. The left part illustrates the structure of \name, including the input layer, the large language model layer and the prediction layer. The upper right part describes the efficient inference process. The lower right part shows the complete solution of \name.}
    \label{fig:model}
\end{figure*}

\section{Methodology}
In this section, we will introduce the complete solution of our proposed \name in detail, including our pathway of ID injection, the tuning strategy of the backbone foundation models, the model structure of \name and the deployment of \name.

\subsection{Overview}
In order to deliver a clearer and more concise schema for better understanding, we provide the whole architecture of our proposed \name solution in Figure~\ref{fig:model}.
As illustrated in the lower right part, our \name solution consists of four stages, which are sequential recommendation model pretraining, LLM instruction tuning, \name model training and \name model deployment.
Specifically, the pretraining of sequential recommendation models and the instruction tuning of LLMs are the preliminaries and these two stages are essentially decoupled from the following stages of \name by only providing sets of parameters as pluggable components.
In the training stage of \name, there are also a small number of parameters of several pluggable components being trained while the entire LLM is frozen and the personalization on the specific dataset is provided by an adapter.
Once being trained, the \name model could be easily deployed to conduct the sequential recommendation task on the given dataset by simply replacing the parameters of the pluggable components.
Through the above stages, we provide the whole solution of \name that can be used to build an industrial-level LLM-based recommender system from scratch.
Please find the detailed description of the whole pipeline of \name solution in Appendix~\ref{sec:A}. 

\subsection{ID Injection}
Although LLMs are powerful in modeling natural language and are able to produce rational responses, they are unaware of the meanings of IDs without textual features and thus are not able to handle pure ID information.
However, the collaborative information contained in the IDs has been proven to be very effective and crucial in personalized recommender systems for a long time.
Therefore, the disability of utilizing ID information strongly limits the practical value of LLMs in recommender systems so far.
Considering the huge number of IDs and the extreme sparsity of collaborative signals, it is especially challenging to incorporate ID information into LLMs.
The existing works~\cite{DBLP:journals/corr/abs-2305-06569,DBLP:journals/corr/abs-2306-11134} have explored various methods to introduce IDs into LLMs for recommendation, including vocabulary expansion, character decomposition, sequential grouping and collaborative clustering.
Nevertheless, these methods are all insufficient to effectively capture the collaborative information.
Meanwhile, the projection of item names or semantic descriptions falls into content-based recommendation without personalization and is only effective when the number of items is quite small with succinct informative textual features.

To address the disability of LLMs to handle ID information, we propose a novel approach by injecting the ID embeddings into the LLMs rather than learning them through the training or tuning process of LLMs.
Specifically, the ID embeddings are obtained by directly extracting from a pretrained sequential recommendation model.
In this paper, we select the SASRec~\cite{DBLP:conf/icdm/KangM18} model to provide ID embeddings considering its effectiveness and generalization. 
It is worth noting that it could be replaced by any other sequential recommendation models for improvement.
The selected sequential recommendation model is pretrained on the given dataset for iterations to get its best performance.
Then the ID embeddings of the items are directly extracted without any modification to be ready for the input of \name, which can be represented as $\mathbf{E} \in \mathbb{R}^{N \times d_s}$.
$N$ and $d_s$ are the number of items and the dimension of ID embeddings, respectively.
Since this sequential recommendation model is pretrained with only sequences of IDs, there is no information on any other features being exposed to the ID embeddings.

\subsection{Backbone Tuning}
The open-sourced LLMs are mainly created for the general purpose of natural language generation to answer any questions from users so that the output formats are diverse and typically wordy.
However, in specific scenarios like recommendation, we expect the generated outputs of LLMs to strictly adhere to the given format usually as simple as a number or a word.
At the same time, it is not desirable for us to get a LLM that is only able to achieve the given task by directly tuning on it.
Hence, we aim to modify the LLMs through instruction tuning~\cite{DBLP:conf/iclr/WeiBZGYLDDL22} on various tasks to enable the LLMs to complete any instructions according to the given template.

Specifically, we choose the LLaMA2-13B~\cite{DBLP:journals/corr/abs-2302-13971} model as the backbone LLM due to its strong power and generalization.
For the instruction tuning process, we follow the settings in Platypus~\cite{DBLP:journals/corr/abs-2308-07317} to apply a Parameter-Efficient Fine-Tuning (PEFT)~\cite{peft} method, LoRA~\cite{DBLP:conf/iclr/HuSWALWWC22}, on the \texttt{gate\_proj}, \texttt{down\_proj} and \texttt{up\_proj} modules of LLaMA2-13B.
Using the PEFT method, only the parameters of the specified modules are trained (about 0.07\% of the total parameters) so that the training efficiency is highly improved.
The initial datasets and more detailed settings can be found at the project page of Platypus\footnote{https://platypus-llm.github.io/}.
The instruction tuning~\cite{DBLP:conf/acl/WangKMLSKH23} is conducted for one epoch using the Alpaca~\cite{alpaca} template.
An example of the Alpaca style prompt is illustrated in Table~\ref{tab:template}, which consists of instruction, input and response parts.

\begin{table}
    \caption{Example of Alpaca template for instruction tuning.}
    \centering
    \begin{tabular}{ll}
        \toprule
        \multicolumn{2}{c}{\textbf{Instruction Input}}\\
        \midrule
        \textbf{\#\#\# Instruction:} &  Please give the maximum common \\
         & subsequence of the following two strings.\\
        \textbf{\#\#\# Input:} & Large Language Model, Language Models \\
        \midrule
        \midrule
        \multicolumn{2}{c}{\textbf{Instruction Output}}\\
        \midrule
        \textbf{\#\#\# Response:} & Language Model \\
        \bottomrule
    \end{tabular}
    \label{tab:template}
\end{table}

\subsection{\name}

\subsubsection{Input Layer}
In this input layer, we aim to inject the IDs into LLMs along with the standard textual prompts.
As illustrated in the left part of Figure~\ref{fig:model}, \name also follows the Alpaca template and the item ID sequences are taken as the input part of the prompt.
From another perspective, each item ID can be seen as a `word' in the LLM but is projected using our proposed ID injection component rather than the lookup table of LLM.
Thus, we obtain the corresponding ID embeddings of the items through the aforementioned ID injection component while the rest of the prompt is projected to word embeddings as the typical LLM inputs.
The final inputs are the concatenation of the word embeddings and the ID embeddings in their original positions.
Since the pretrained ID embeddings usually have a much lower dimension (e.g., 64 in our pretrained SASRec) than the word embeddings in LLM (e.g., 5120 in LLaMA2-13B), we employ a linear projection to convert the ID embeddings into the same dimension with the word embeddings.
As mentioned above, we do not want the LLMs to encounter the `catastrophic forgetting' phenomenon during the training on the recommendation task. 
Certainly, we also do not want the learned collaborative information to be destroyed either.
Therefore, only the linear projection component is trainable, which is as small as a weight matrix $\mathbf{W}^{Input} \in \mathbb{R}^{d_s \times d_k}$.
$d_s$ and $d_k$ represent the dimensions of the ID embeddings in pretrained sequential recommendation model and the word embeddings in LLM, respectively.

\subsubsection{Large Language Model Layer}
The large language model layer employs the instruction-tuned LLaMA2-13B, which is the merge of the original LLaMA2-13B and the LoRA adapter parameters for instruction tuning.
Similar to the LLM instruction tuning stage, we also introduce an additional LoRA adapter on the \texttt{gate\_proj}, \texttt{down\_proj} and \texttt{up\_proj} modules to model the personalization of the given recommendation task.

\subsubsection{Prediction Layer}
The prediction layer receives the output of the LLM and makes the predictions for recommendation.
Existing LLM-based recommendation approaches mainly define the recommendation task as an open-domain natural language generation task, which is in line with the purpose of LLMs themselves.
However, due to the characteristics of the task definition, they often generate out-of-range results and are only able to generate one recommendation result each time.
Such a pathway impairs the reliability of the recommendation results and limits the recommendation efficiency.
Therefore, these approaches can only work in the ranking stages (usually less than one hundred items) but are impractical in the matching stages (usually more than thousands of items).

Unlike these existing approaches that directly adopt the task definition of LLMs without modification, we stick to their underlying definition of modeling joint probability distribution for generation.
That is to say, we can compute the prediction results for a given sequence over all the candidates in each forward process.
In order to achieve this goal, we dive deep into the structure of LLMs and propose to employ an item linear projection to replace the original prediction layer in LLMs via a weight matrix $\mathbf{W}^{Output} \in \mathbb{R}^{d_k \times N}$ where $N$ denotes the total number of candidate items.
Then, the predictions of a given sequence could be represented as a $N$-dimensional vector $\mathbf{\hat{y}} \in \mathbb{R}^{N}$.
In the training stage, we adopt the cross entropy loss between predictions and ground-truth next items as the learning objective of \name as follows:
\begin{equation}\nonumber\label{eq:loss}
    \mathcal{L} = -\sum_{i=1}^N y_i \mathrm{log}(\hat{y}_i)
\end{equation}

\subsection{Inference \& Deployment}
The inference time is also an important and bothering problem for the application of LLMs.
Although we cannot improve the performance of LLMs themselves, our proposed \name solution can ensure that the overall inference time of the recommendation task is as close as that of a vanilla LLM.
Review the structure of our proposed \name, the additional components are all so small compared with the backbone LLM that they will only cost a little more time.
More than that, the time-consuming \texttt{softmax} operation in the training stage is also unnecessary. 
Therefore, the inference process can be further simplified to the nearest neighbor search between the output of LLM and the vectors in the item linear projection component as shown in the upper right part of Figure~\ref{fig:model}.

After the training stage is completed, our proposed \name can be deployed in a very lightweight manner.
The backbone LLM is one-size-fits-all for all the tasks including sequential recommendation so that the instruction tuning only needs to be done once and is shared across all the downstream tasks.
For each coming sequential recommendation dataset, only the ID embeddings $\mathbf{E}$, the linear projection in the input layer $\mathbf{W}^{Input}$, LoRA weights $\Theta$ and the item linear projection $\mathbf{W}^{Output}$ are required to be trained and stored for deployment.
Compared with the billions of parameters in the LLMs, these parameters are as tiny as about 1\%.
Meanwhile, all these components are completely pluggable so that the recommender system can quickly adapt to a specific dataset by simply replacing these pluggable components.

%% file: src/4-exper.tex
\begin{table}
    \caption{Statistics of the datasets.}
    \centering
    \begin{tabular}{lrrrr}
        \toprule
        \textbf{Dataset} & \textbf{\# Users} & \textbf{\# Items} & \textbf{\# Actions} & \textbf{Sparsity} \\
        \midrule
        Beauty & 22,363 & 12,101 & 198,502 & 99.93\% \\
        Sports & 25,598 & 18,357 & 296,337 & 99.95\% \\
        Toys & 19,412 & 11,924 & 167,597 & 99.93\% \\
        Yelp & 30,431 & 20,033 & 316,354 & 99.95\% \\
        \bottomrule
    \end{tabular}
    \label{tab:datasets}
\end{table}

\begin{table*}
    \caption{Performance comparison of different methods. The best performance is highlighted in bold while the second best performance is underlined. The last column indicates the improvements over the best baseline models and all the results of \name are statistically significant with p < 0.01 compared to the best baseline models.}
    \centering
    \scalebox{0.95}
    {
    \begin{tabular}{ll|cc|ccc|cccc|c|r}
        \toprule
        Dataset & Metric & POP & BPR & GRU4Rec & Caser & SASRec & BERT4Rec & S$^3$-Rec & CL4SRec & ICLRec & \textbf{\name} & Improv. \\
        \midrule
        \multirow{6}*{Beauty} 
        & HR@5 & 0.0072 & 0.0120 & 0.0164 & 0.0251 & 0.0333 & 0.0193 & 0.0327 & 0.0407 & \underline{0.0436} & \textbf{0.0525} & 20.41\% \\
        & HR@10 & 0.0114 & 0.0361 & 0.0289 & 0.0418 & 0.0581 & 0.0401 & 0.0591 & 0.0626 & \underline{0.0653} & \textbf{0.0758} & 16.08\% \\
        & HR@20 & 0.0195 & 0.0589 & 0.0478 & 0.0643 & 0.0915 & 0.0596 & 0.0898 & 0.0957 & \underline{0.0974} & \textbf{0.1071} & 9.96\% \\
        & nDCG@5 & 0.0040 & 0.0065 & 0.0086 & 0.0127 & 0.0179 & 0.0187 & 0.0175 & 0.0223 & \underline{0.0240} & \textbf{0.0360} & 50.00\% \\
        & nDCG@10 & 0.0053 & 0.0122 & 0.0142 & 0.0193 & 0.0258 & 0.0254 & 0.0268 & 0.0317 & \underline{0.0338} & \textbf{0.0435} & 28.70\% \\
        & nDCG@20 & 0.0073 & 0.0179 & 0.0169 & 0.0258 & 0.0342 & 0.0361 & 0.0370 & 0.0396 & \underline{0.0416} & \textbf{0.0514} & 23.56\% \\
        \midrule
        \multirow{6}*{Sports} 
        & HR@5 & 0.0055 & 0.0092 & 0.0137 & 0.0139 & 0.0170 & 0.0176 & 0.0157 & 0.0217 & \underline{0.0238} & \textbf{0.0281} & 18.07\% \\
        & HR@10 & 0.0090 & 0.0188 & 0.0274 & 0.0231 & 0.0289 & 0.0326 & 0.0265 & 0.0374 & \underline{0.0393} & \textbf{0.0410} & 4.33\% \\
        & HR@20 & 0.0149 & 0.0258 & 0.0438 & 0.0389 & 0.0477 & 0.0493 & 0.0460 & \underline{0.0582} & 0.0553 & \textbf{0.0626} & 7.56\% \\
        & nDCG@5 & 0.0040 & 0.0053 & 0.0096 & 0.0085 & 0.0091 & 0.0105 & 0.0098 & 0.0129 & \underline{0.0152} & \textbf{0.0196} & 28.95\% \\
        & nDCG@10 & 0.0051 & 0.0083 & 0.0137 & 0.0126 & 0.0129 & 0.0153 & 0.0135 & 0.0184 & \underline{0.0212} & \textbf{0.0237} & 11.79\% \\
        & nDCG@20 & 0.0066 & 0.0121 & 0.0171 & 0.0166 & 0.0177 & 0.0195 & 0.0182 & 0.0239 & \underline{0.0250} & \textbf{0.0291} & 16.40\% \\
        \midrule
        \multirow{6}*{Toys} 
        & HR@5 & 0.0064 & 0.0120 & 0.0097 & 0.0166 & 0.0445 & 0.0274 & 0.0492 & 0.0484 & \underline{0.0509} & \textbf{0.0566} & 11.20\% \\
        & HR@10 & 0.0079 & 0.0211 & 0.0196 & 0.0281 & 0.0698 & 0.0460 & 0.0698 & 0.0706 & \underline{0.0725} & \textbf{0.0798} & 10.07\% \\
        & HR@20 & 0.0108 & 0.0312 & 0.0301 & 0.0420 & 0.0999 & 0.0688 & 0.0962 & 0.0984 & \underline{0.1018} & \textbf{0.1107} & 8.74\% \\
        & nDCG@5 & 0.0037 & 0.0082 & 0.0059 & 0.0107 & 0.0236 & 0.0174 & 0.0342 & 0.0327 & \underline{0.0350} & \textbf{0.0405} & 15.71\% \\
        & nDCG@10 & 0.0057 & 0.0120 & 0.0098 & 0.0151 & 0.0318 & 0.0230 & 0.0375 & 0.0404 & \underline{0.0423} & \textbf{0.0479} & 13.24\% \\
        & nDCG@20 & 0.0062 & 0.0136 & 0.0116 & 0.0179 & 0.0394 & 0.0291 & 0.0431 & 0.0466 & \underline{0.0493} & \textbf{0.0557} & 12.98\% \\
        \midrule
        \multirow{6}*{Yelp} 
        & HR@5 & 0.0056 & 0.0127 & 0.0152 & 0.0142 & 0.0161 & 0.0186 & 0.0173 & 0.0216 & \underline{0.0240} & \textbf{0.0266} & 10.83\% \\
        & HR@10 & 0.0083 & 0.0245 & 0.0263 & 0.0252 & 0.0265 & 0.0291 & 0.0282 & 0.0352 & \underline{0.0381} & \textbf{0.0418} & 9.71\% \\
        & HR@20 & 0.0120 & 0.0346 & 0.0371 & 0.0406 & 0.0443 & 0.0564 & 0.0538 & 0.0585 & \underline{0.0630} & \textbf{0.0675} & 7.14\% \\
        & nDCG@5 & 0.0036 & 0.0076 & 0.0104 & 0.0096 & 0.0102 & 0.0121 & 0.0114 & 0.0130 & \underline{0.0150} & \textbf{0.0189} & 26.00\% \\
        & nDCG@10 & 0.0043 & 0.0119 & 0.0137 & 0.0129 & 0.0134 & 0.0171 & 0.0163 & 0.0185 & \underline{0.0203} & \textbf{0.0238} & 17.24\% \\
        & nDCG@20 & 0.0056 & 0.0143 & 0.0145 & 0.0156 & 0.0179 & 0.0223 & 0.0201 & 0.0235 & \underline{0.0256} & \textbf{0.0297} & 16.02\% \\
        \bottomrule
    \end{tabular}
    }
    \label{tab:results}
\end{table*}

\section{Experiments}
In this section, we will evaluate our proposed \name on several real-world datasets with a selected set of widely-used baseline methods in sequential recommendation.
Meanwhile, we will also present the ablation study, robustness analysis, efficiency analyses and discussions in order to answer the following questions:

\begin{itemize}[leftmargin=*]
    \item \textbf{RQ1:} How does \name compare with traditional sequential recommendation models on performance?
    \item \textbf{RQ2:} How do the injected IDs and the large language model affect the performance of \name?
    \item \textbf{RQ3:} How effective is \name in leveraging collaborative information to alleviate the data sparsity problem?
    \item \textbf{RQ4:} How efficient is \name on 
    both inference time and storage space in deployment?
    \item \textbf{RQ5:} How extensible is \name for new items in industrial applications?
\end{itemize}

\subsection{Experimental Settings}

\subsubsection{Datasets}

To evaluate the effectiveness of \name, we conduct the experiments on four widely-used real-world datasets.
Specifically, \textit{Beauty}, \textit{Sports} and \textit{Toys} are the datasets of sub-categories `Beauty', `Sports and Outdoors' and `Toys and Games' in the Amazon review data~\cite{DBLP:conf/sigir/McAuleyTSH15}.
\textit{Yelp} is a popular platform and the dataset is widely-used in various recommendation tasks.
Here, we only utilize the data after January 1st, 2019.
The statistics of the datasets are shown in Table~\ref{tab:datasets}.

For sequential recommendation, the interaction sequences of users are sorted by timestamps in ascending order.
Following the previous works~\cite{DBLP:conf/www/RendleFS10,DBLP:conf/cikm/SunLWPLOJ19}, we apply the 5-core settings to filter the unpopular items with fewer than 5 interactions to ensure robust evaluation.

\subsubsection{Evaluation Metrics}
To avoid selection bias and provide more reliable results, we evaluate the performance of predictions on the whole item set, which is effective in evaluating the matching ability.
Following previous works~\cite{DBLP:conf/icdm/KangM18,DBLP:conf/sigir/RenLLZWDW20}, we also apply the \textit{leave-one-out} strategy.
For each interaction sequence of users, the last item is taken as test data, the second last one is taken as validation data and the remaining sequence is used for training.
As for the evaluation metrics, we choose two types of widely-used metrics, which are top-k Hit Ratio (HR@k) and top-k normalized Discounted Cumulative Gain (nDCG@k) with k = \{5, 10, 20\}.
These metrics are averaged over all the users for report.

Besides, we also perform an additional evaluation with negative sampling in the Appendix~\ref{sec:C} according to~\cite{DBLP:conf/cikm/ZhouWZZWZWW20,DBLP:conf/www/ZhouYZW22}.
Specifically, 99 negative items are randomly sampled for each positive item.
In this setting, we are able to evaluate the ranking ability of \name.

\subsubsection{Baseline Methods}
The baseline methods chosen for comparison can be split into three categories: non-sequential methods, traditional sequential methods and self-supervised sequential methods.

For non-sequential methods, we have:
\begin{itemize}[leftmargin=*]
    \item \textbf{POP} is a heuristic method that directly ranks the items using their popularity defined as the interaction numbers.
    \item \textbf{BPR}~\cite{DBLP:conf/uai/RendleFGS09} utilizes the Bayesian Personalized Ranking (BPR) loss to optimize the matrix factorization (MF)~\cite{DBLP:journals/computer/KorenBV09} model for characterizing the pair-wise interactions.
\end{itemize}

For traditional sequential methods, we have:
\begin{itemize}[leftmargin=*]
    \item \textbf{GRU4Rec}~\cite{DBLP:journals/corr/HidasiKBT15} implements the GRU recurrent neural network for sequential modeling and then makes predictions for recommendation.
    \item \textbf{Caser}~\cite{DBLP:conf/wsdm/TangW18} integrates both horizontal and vertical convolutional operations to better capture the high-order interactions within item sequences for recommendation.
    \item \textbf{SASRec}~\cite{DBLP:conf/icdm/KangM18} is a self-attentive sequential recommendation model with multi-head self-attention to model the complex sequential information.
\end{itemize}    

For self-supervised sequential methods, we have:
\begin{itemize}[leftmargin=*]
    \item \textbf{BERT4Rec}~\cite{DBLP:conf/cikm/SunLWPLOJ19} employs the Cloze~\cite{Taylor1953ClozePA} objective rather than the next-item prediction for sequential recommendation in a pretraining-tuning manner.
    \item \textbf{S$^3$-Rec}~\cite{DBLP:conf/cikm/ZhouWZZWZWW20} applies contrastive learning to capture the correlations among items, sub-sequences and attributes. Specifically, we take the variant with only Mask Item Prediction (MIP) objective.
    \item \textbf{CL4SRec}~\cite{DBLP:conf/icde/XieSLWGZDC22} incorporates the contrastive learning with the transformer-based sequential recommendation model to obtain more robust results.
    \item \textbf{ICLRec}~\cite{DBLP:conf/www/ChenLLMX22} leverages a latent intent variable to learn the users’ intent distribution from unlabeled item sequences to improve the transformer-based sequential recommendation model.
\end{itemize}

Note that our proposed \name only utilizes the ID information and no other features are exposed to \name during the training and inference.
Therefore, those methods that either implement data augmentation techniques~\cite{DBLP:conf/www/ZhouYZW22,DBLP:conf/recsys/ZhouGXYHKW023} or incorporate other features~\cite{DBLP:conf/wsdm/QiuHYW22,DBLP:conf/sigir/DuYZZ00LS23} are orthogonal to \name and are thus excluded for fair comparison.

\subsubsection{Implementation Details}
For GRU4Rec, Caser, BERT4Rec, S$^3$-Rec and ICLRec, we implement them using the public resources released by their authors.
For other models, we implement them using PyTorch 2.0.1.
The embedding dimension is set to 64 and the maximum sequence length is set to 50 for all the models on all the datasets.
The model parameters are initialized with Xavier initialization and are optimized using Adam~\cite{DBLP:journals/corr/KingmaB14}.

For our proposed \name, we obtain the LLaMA2-13B using HuggingFace\footnote{https://huggingface.co/meta-llama/Llama-2-13b} and conduct instruction tuning for one epoch.
The ID embeddings are directly extracted from the pretrained SASRec model without any modification.
To obtain better performance, we perform the grid search of the training configurations using the validation set.
Specifically, we aim to find a better combination of learning rate, training epochs and LoRA modules.
The best combinations for all the datasets are listed in Appendix~\ref{sec:B} due to limited space.
All the experiments are implemented using 8 NVIDIA Tesla A800 GPUs.
The implementation code is available online\footnote{https://github.com/HestiaSky/E4SRec/} and will be accessible to the public for ease of reproducibility.

\subsection{Main Results (RQ1)}

The performance comparison of our proposed \name with other traditional sequential recommendation models is shown in Table~\ref{tab:results}.
Here we have the following observations:

\begin{itemize}[leftmargin=*]
    \item Our proposed \name can significantly outperform all the baseline methods on all four datasets thanks to the powerful ability of LLM. The relative improvements on performance over the best baseline methods are about 12\% on HR@k and 21\% on nDCG@k, which fully demonstrate the effectiveness of \name.
    \item Generally, the performance gains are greater on nDCG@k metrics than on HR@k metrics and are greater on smaller k values. This indicates that our proposed \name can capture the users' preference more accurately and thus generate more reliable recommendation results.
    \item The non-sequential methods are much worse than sequential methods due to the disadvantage of modeling sequential information. The dramatically bad performance of POP also indicates the important and crucial role of personalization in these datasets.
    \item The self-supervised sequential methods are usually stronger than the traditional sequential methods. This phenomenon shows the effectiveness of introducing self-supervised learning to provide additional training signals in improving the sequential recommendation performance.
    \item CL4SRec and ICLRec achieve much better performance with the help of contrastive learning in improving the robustness of item representations. However, contrastive learning applies data augmentation in a certain extent by introducing augmented sequences, which may be a little bit unfair to the other methods. Therefore, SASRec is still a very strong method compared with them and is more suitable for reference. In such a situation, our proposed \name will have an even more significant performance gain of up to 115\%.
\end{itemize}

\begin{figure*}
    \centering
    \includegraphics[width=\linewidth]{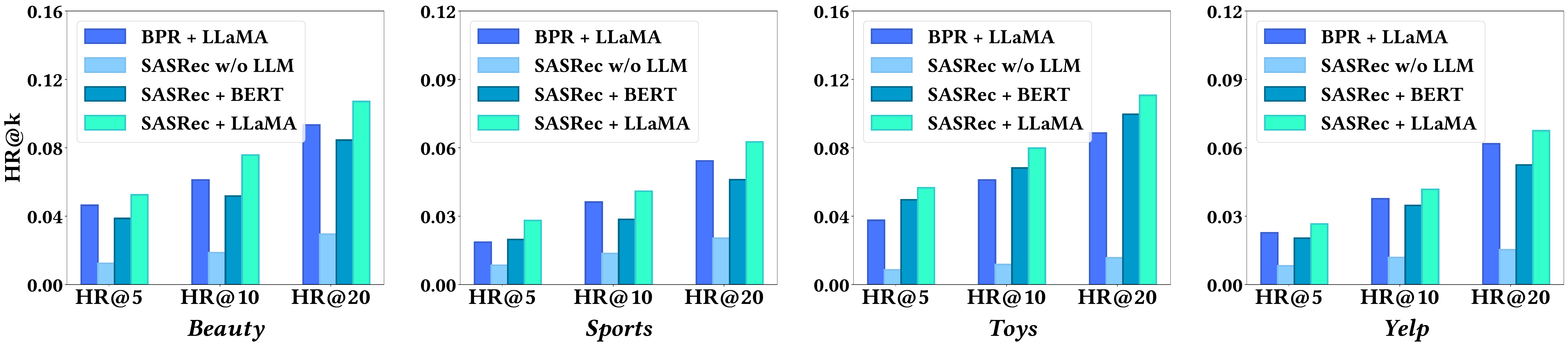}
    \caption{Ablation study of ID injection and LLM.}
    \label{fig:ablation}
\end{figure*}

\subsection{Ablation Study (RQ2)}

In order to explore the impacts of the injected IDs and the LLMs on the overall performance of \name, we perform an ablation study by changing the injected IDs and the LLMs to design and then compare the following four variants:
\begin{itemize}[leftmargin=*]
    \item \textbf{BPR + LLaMA}: This variant employs the BPR model to provide item ID embeddings and the LLaMA2-13B model.
    \item \textbf{SASRec w/o LLM}: This is a LLM-free variant that utilizes the SASRec as the sequential model for ID injection. Specifically, it has no LLM and only uses linear projections for prediction.
    \item \textbf{SASRec + BERT}: A BERT~\cite{DBLP:conf/naacl/DevlinCLT19} model is utilized in the LLM layer. Here we use \texttt{BERT-base-uncased} with 110M parameters.
    \item \textbf{SASRec + LLaMA}: It is the basic version of our proposed \name for reference with SASRec for ID injection and LLaMA2-13B as the LLM.
\end{itemize}

From the performance comparison of HR@k metrics in Figure~\ref{fig:ablation}, we have the following observations:
\begin{itemize}[leftmargin=*]
    \item The qualities of item ID embeddings do have effects on the performance. The performance of BPR + LLaMA is slightly worse than SASRec + LLaMA. Although BPR is unable to capture sequential information and has poor performance, the performance of BPR + LLaMA is still satisfying. This phenomenon may indicate that the main effect of ID injection is to provide collaborative information while the LLM is already sufficient to well capture sequential information during the training stage.
    \item The extremely poor performance of SASRec w/o LLM demonstrates that the LLMs are necessary to conduct recommendation.
    \item The inherent ability of the LLMs also has a significant impact on overall performance. Although BERT is already a very powerful language model, the performance gap is still significant compared to LLaMA. This also implies the effectiveness of our proposed \name solution in leveraging the capabilities and tapping into the potentials of LLMs.
\end{itemize}

\subsection{Robustness Analysis (RQ3)}

\begin{figure}[t]
    \centering
    \includegraphics[width=\linewidth]{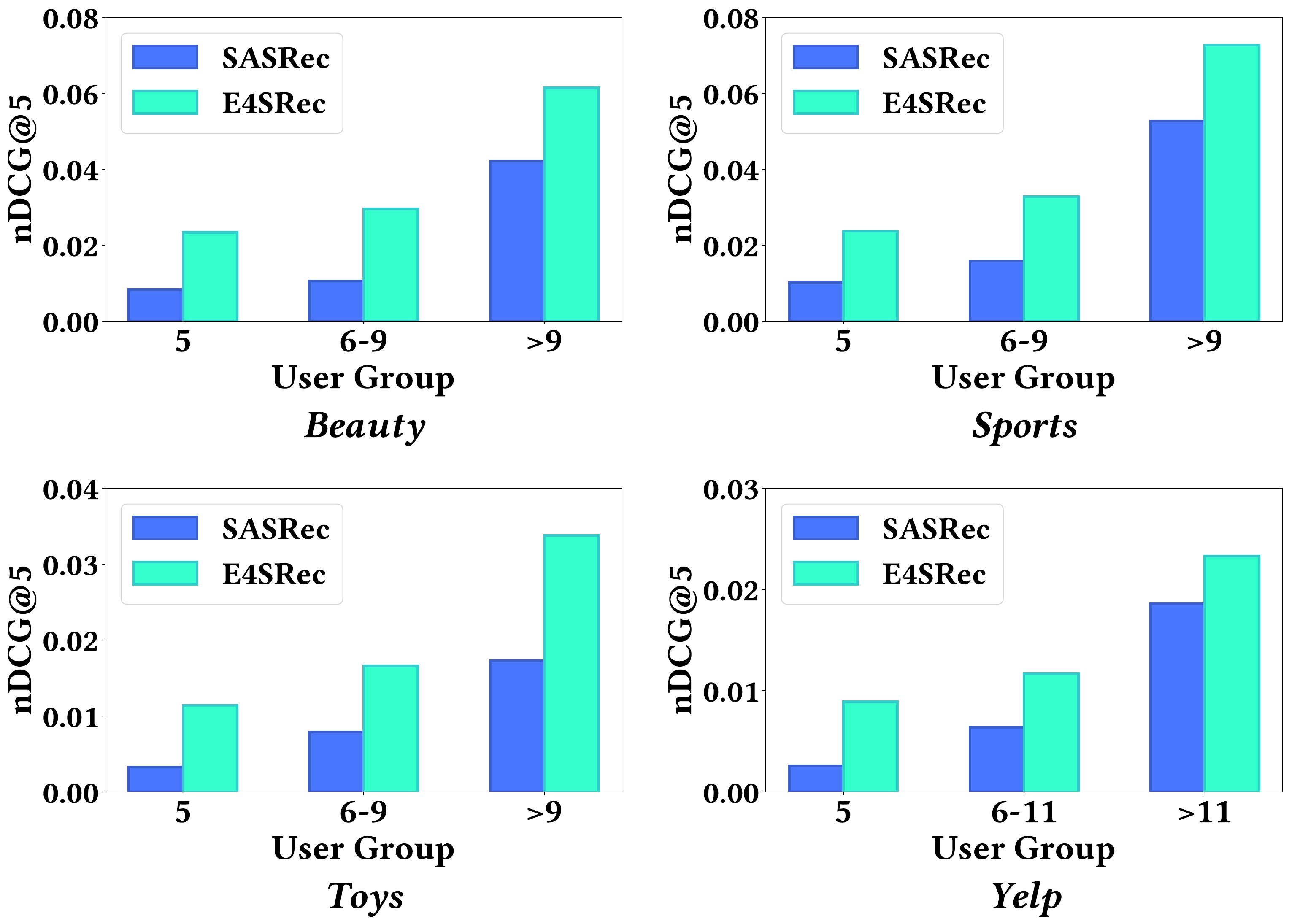}
    \caption{Performance comparison on different user groups with respect to the length of interaction sequences.}
    \label{fig:long-tail}
\end{figure}

Data sparsity problem is a common issue of recommender systems that defects the performance in applications.
For example, since most of the users only have limited interactions, the user cold-start problem is typically severe and thus harms the user experience. 
Traditional recommendation methods alleviate the data sparsity problems by jointly leveraging the data from other users and items via collaborative filtering.
To verify the effectiveness of our proposed \name in incorporating collaborative information, we compared it with SASRec on the robustness with different sparsity levels' data as shown in Figure~\ref{fig:long-tail}.
Specifically, we split the users into three groups based on their number of interactions.
Based on the results, we have observations as follows:
\begin{itemize}[leftmargin=*]
    \item \name exceeds SASRec on all three user groups with significant margins, which demonstrates the effectiveness in modeling collaborative information.
    \item The convincing results on the sparse user group with merely 5 interactions, in which there is only one training instance per user, also prove the robustness of our proposed solution on extremely sparse data.
    \item The performance gains are more significant on sparser user groups than denser ones. Such a phenomenon may be originated from the few-/zero-shot learning ability of LLMs to enable \name to achieve satisfying performance with insufficient data that traditional recommendation methods can hardly learn from.
\end{itemize}

\subsection{Efficiency Analysis (RQ4)}

In addition to being effective in sequential recommendation, our proposed \name solution is also very efficient in both time and space.
To support this statement, we provide a comparison of inference time and storage space between the backbone LLM, LLaMA2-13B and \name on 1 Nvidia Tesla A800 GPU as illustrated in Table~\ref{tab:efficiency}.
As mentioned above, \name introduces only a tiny pluggable set of parameters to the backbone LLMs and doesn't change the overall data flow too much.
Therefore, the inference time of \name is very close to the backbone LLaMA2-13B.
Considering the ability to generate results without the need for multiple times generation, our \name is obviously more efficient by orders of magnitude on the actual inference time than the existing solutions which directly apply the task formulation of natural language generation.
Meanwhile, the additional parameters in \name are as tiny as 1\% of the total parameters of the backbone LLM.
Since this set of additional parameters is pluggable to have no effect on the backbone LLM, \name can be stored and deployed with only one shared backbone LLM and multiple independent pluggable components for specific datasets.
In this manner, there is only 2 times of space needed than the backbone LLM to store 100 \name models, which is very economical in industrial applications.

\begin{table}[t]
    \caption{Comparisons of inference time and storage space between LLaMA2-13B and \name.}
    \centering
    \begin{tabular}{c|cc}
        \toprule
        \textbf{Model} & \textbf{Inference Time} & \textbf{Storage Space}\\
        \midrule
        LLaMA2-13B & 0.21s/Instance & 12.12GB\\
        \name & 0.23s/Instance & 12.23GB\\
        \bottomrule
    \end{tabular}
    \label{tab:efficiency}
\end{table}

\subsection{Discussions (RQ5)}

In industrial applications, there are many new items emerging into the recommender systems every day.
Apparently, retraining the recommendation model for each coming item is not practical so the ability to extend to new items is very crucial.
Our \name solution guarantees that the pluggable components are independent of IDs, which means adding a new item in the dataset only requires adding a new row in the linear projections without the need to retrain the entire model.
Therefore, our proposed \name solution is considerably extensible.

%% file: src/2-relatedworks.tex
\section{Related Works}

\subsection{Sequential Recommendation}
Sequential recommendation is an important task in personalized recommender system which aims to capture the users' preference using their historical behavior sequences.
Early works of sequential recommendation mainly lie in the pattern of Markov Chains~\cite{Rendle2010FactorizationM,DBLP:conf/icdm/HeM16} that uses an item-item
transition pattern to directly predict the next item with the previous items.
For example, FPMC~\cite{DBLP:conf/www/RendleFS10} uses Matrix Factorization (MF) to model the users' preference and utilizes Markov Chains to capture the sequential patterns for making prediction.
With the booming development of deep learning, many sequential recommendation models with deep neural networks emerged.
Caser~\cite{DBLP:conf/wsdm/TangW18} employs Convolutional Neural Network (CNN) and GRU4Rec~\cite{DBLP:journals/corr/HidasiKBT15} utilizes Recurrent Neural Network (RNN) to capture the high-order interactions within the item sequences for sequential modeling.
In recent years, more powerful models are proposed with more advanced architectures to better leverage sequential information.
SASRec~\cite{DBLP:conf/icdm/KangM18} takes advantage of the multi-head self-attention to attentively model sequential information in an unidirectional manner.
BERT4Rec~\cite{DBLP:conf/cikm/SunLWPLOJ19} improves such a manner by employing the Cloze~\cite{Taylor1953ClozePA} objective to predict the masked item to leverage the bidirectional information.
S$^3$-Rec~\cite{DBLP:conf/cikm/ZhouWZZWZWW20} introduces contrastive learning to fuse the information of distinct items, sub-sequences and attributes.
Similarly, CL4SRec~\cite{DBLP:conf/icde/XieSLWGZDC22} and ICLRec~\cite{DBLP:conf/www/ChenLLMX22} both apply contrastive learning to better capture the users' preference with sequential information.
There are also many other approaches that aim to incorporate other features, e.g. DuoRec~\cite{DBLP:conf/wsdm/QiuHYW22} and EMKD~\cite{DBLP:conf/sigir/DuYZZ00LS23}, or employ data augmentation, e.g. FMLP-Rec~\cite{DBLP:conf/www/ZhouYZW22} and 
ECL-SR~\cite{DBLP:conf/recsys/ZhouGXYHKW023} for further improvements.

Nevertheless, traditional sequential recommendation models are usually limited on performance for the limitations on the model scale and may lead to sub-optimal prediction.
Unlike them, our proposed \name takes advantages of LLMs to achieve more advanced performance and empower the generative recommendation, which can better comprehend human intentions and generate more human-like language responses.

\subsection{LLMs for Recommendation}
Large Language Models (LLMs) have been proven to be very powerful in natural language processing and their strong power has encouraged researchers to make efforts to apply LLM for recommendation~\cite{DBLP:journals/corr/abs-2307-02046}.
Early approaches view LLMs as feature extractors to generate knowledge-aware embeddings for recommendation.
U-BERT~\cite{DBLP:conf/aaai/QiuWG021} proposes a pretraining-tuning framework to learn users’ representations and conduct user modeling.
UserBERT~\cite{DBLP:conf/sigir/WuWQH22} employs two self-supervision tasks on unlabeled behavior data to empower user modeling.
With the emergence of generative LLMs like GPT, LLM-based recommendation has also shifted towards generative recommendation.
These methods translate recommendation tasks as natural language tasks to directly generate the recommendation results~\cite{DBLP:journals/corr/abs-2305-19860}.
At first, most approaches focus on using prompting~\cite{DBLP:journals/corr/abs-2304-09542,DBLP:journals/corr/abs-2303-14524} or in-context learning~\cite{DBLP:journals/corr/abs-2304-10149,DBLP:conf/recsys/DaiSZYSXS0X23} to adapt LLMs for recommendation.
However, these approaches fail to surpass the traditional recommendation models trained specifically for a given task on specific data.
Therefore, many efforts are made to align the LLMs to recommendation by further fine-tuning recently.
P5~\cite{DBLP:conf/recsys/Geng0FGZ22} first proposes a unified framework to integrate five recommendation tasks via fine-tuning on FLAN-T5~\cite{DBLP:journals/jmlr/RaffelSRLNMZLL20}.
Following it, InstructRec~\cite{DBLP:journals/corr/abs-2305-07001} adapts FLAN-T5 model to several downstream recommendation tasks by instruction tuning with more diverse texts.
TALLRec~\cite{DBLP:conf/recsys/BaoZZWF023} aligns the LLaMA model to the binary recommendation task by two stages of instruction tuning in few-shot scenario.
GenRec~\cite{DBLP:journals/corr/abs-2307-00457} directly conducts instruction tuning on the LLaMA model with plain texts to achieve generative recommendation.

However, all the above methods are essentially content-based recommendation and require rich semantic features to achieve satisfying performance.
Therefore, they fail to handle the IDs and are unable to leverage collaborative information.
Compared with them, our proposed \name solution is more effective in handling IDs, more efficient on both time and space perspective and more extensible to fulfill real-world needs in application.

%% file: src/5-conclusion.tex
\section{Conclusion}
Existing approaches of LLM for recommendation face challenges in handling IDs, efficiency, extensiblility and thus are not able to fulfill the requirements of real-world applications.
In this paper, we propose a novel \name solution, which is elegant, effective, efficient and extensible to apply LLMs for sequential recommendation.
Specifically, we introduce an elegant way to address the issue of handling IDs by injecting item ID embeddings into the LLM.
Meanwhile, with the help of a modified prediction layer, we effectively solve the challenging out-of-range problem of generated results to ensure their legality and efficiently generate the predictions over all the candidates at once.
The design of pluggable components in \name enables the model to be trained and deployed in a lightweight manner.
The extensive experiments on four popular read-world datasets fully demonstrate the effectiveness and superiority of our proposed \name solution.

We hope our \name solution will contribute to the research in applying LLM for recommender systems.
Looking forward, we will pivot towards crafting elegant solutions for other recommendation tasks, such as CTR prediction, and continually pushing the frontiers of generative recommendations.

%% file: src/6-appendix.tex
\appendix
\newpage

\RestyleAlgo{ruled}
\begin{algorithm}[t]
\SetKwInOut{Input}{Input}
\SetKwInOut{Output}{Output}
\SetKwProg{Fn}{Function}{:}{\KwRet $\mathbf{T}$}
\caption{Pipeline of \name Solution.}\label{alg:solution}

\Input{recommendation dataset $\mathcal{D}$, instruction dataset $\mathcal{D}_{Ins}$ and pretrained LLM $\mathcal{M}_{LLM}$}
\Output{E4SRec model $\mathcal{M}$}

\tcp{Preliminary}
Train a sequential recommendation model $\mathcal{M}_{Seq}$ on $\mathcal{D}$.

Extract item ID embeddings $\mathbf{E}$.

Instruction tune the LLM model $\mathcal{M}_{LLM}$ on $\mathcal{D}_{Ins}$.

\tcp{Model Training}
Initialize pluggable components $\Theta$, including input linear projection, LoRA weights and item linear projection.

\For{$T \gets 0$ \KwTo $T_{max}$ iterations}{
    Sample an instance for training.
    
    \tcp{ID Injection}
    Obtain the corresponding ID embeddings from $\mathbf{E}$;
    Project ID embeddings to the same dimension with $\mathcal{M}_{LLM}$ using input linear projection.
    
    \tcp{Prediction}
    Feed ID embeddings and prompt to $\mathcal{M}_{LLM}$ for output;
    Predict candidate items using item linear projection.
    
    \tcp{Update Parameters}
    Compute the cross-entropy loss $\mathcal{L}$ via Equation~\ref{eq:loss};
    Update $\Theta$ using $\mathcal{L}$.
}

\tcp{Model Deployment}
Deploy backbone model $\mathcal{M}_{LLM}$.

Deploy E4SRec model for dataset $\mathcal{D}$ with $\mathcal{M} = \mathcal{M}_{LLM} \gets \mathbf{E}, \Theta$.

\end{algorithm}

\begin{table}[t]
    \caption{Training configuration on the four datasets.}
    \centering
    \begin{tabular}{l|cccc}
        \toprule
        \textbf{Parameter} & \textbf{Beauty} & \textbf{Sports} & \textbf{Toys} & \textbf{Yelp} \\
        \midrule
        Training Epochs & 3 & 3 & 2 & 5\\
        Learning Rate & 3e-4 & 2e-4 & 2e-4 & 3e-4\\
        Batch Size & \multicolumn{4}{c}{16}\\
        LoRA Rank & \multicolumn{4}{c}{16}\\
        LoRA Alpha & \multicolumn{4}{c}{16}\\
        LoRA Dropout & \multicolumn{4}{c}{0.05}\\
        LoRA Modules & \multicolumn{4}{c}{[\texttt{gate\_proj}, \texttt{down\_proj}, \texttt{up\_proj}]}\\
        Learning Rate Scheduler & \multicolumn{4}{c}{Cosine Scheduler}\\
        Weight Decay & \multicolumn{4}{c}{0.1}\\
        Warmup Steps & 100 & 200 & 100 & 300\\
        \bottomrule
    \end{tabular}
    \label{tab:config}
\end{table}

\begin{table*}
    \caption{Performance comparison of different methods with sampled negative items. The best performance is highlighted in bold while the second best performance is underlined. The last column indicates the improvements over the best baseline models and all the results of \name are statistically significant with p < 0.01 compared to the best baseline models.}
    \centering
    \begin{tabular}{ll|cccccc|cr}
        \toprule
        Dataset & Metric & POP & BPR & GRU4Rec & Caser & SASRec & BERT4Rec & \textbf{\name} & Improv. \\
        \midrule
        \multirow{6}*{Beauty} 
        & HR@1 & 0.0678 & 0.0405 & 0.1337 & 0.1337 & \underline{0.1870} & 0.1531 & \textbf{0.2274} & 21.60\% \\
        & HR@5 & 0.2105 & 0.1461 & 0.3125 & 0.3032 & \underline{0.3741} & 0.3640 & \textbf{0.4088} & 9.28\% \\
        & nDCG@5 & 0.1391 & 0.0934 & 0.2268 & 0.2219 & \underline{0.2848} & 0.2622 & \textbf{0.3221} & 13.10\% \\
        & HR@10 & 0.3386 & 0.2311 & 0.4106 & 0.3942 & 0.4696 & \underline{0.4739} & \textbf{0.5068} & 6.94\% \\
        & nDCG@10 & 0.1803 & 0.1207 & 0.2584 & 0.2512 & \underline{0.3156} & 0.2975 & \textbf{0.3503} & 10.99\% \\
        & MRR & 0.1558 & 0.1096 & 0.2308 & 0.2263 & \underline{0.2852} & 0.2614 & \textbf{0.3182} & 11.57\% \\
        \midrule
        \multirow{6}*{Sports} 
        & HR@1 & 0.0763 & 0.0489 & 0.1160 & 0.1135 & \underline{0.1455} & 0.1255 & \textbf{0.1732} & 19.04\% \\
        & HR@5 & 0.2293 & 0.1603 & 0.3055 & 0.2866 & \underline{0.3466} & 0.3375 & \textbf{0.3721} & 7.36\% \\
        & nDCG@5 & 0.1538 & 0.1048 & 0.2126 & 0.2020 & \underline{0.2497} & 0.2341 & \textbf{0.2701} & 8.17\% \\
        & HR@10 & 0.3423 & 0.2491 & 0.4299 & 0.4014 & 0.4622 & \underline{0.4722} & \textbf{0.4821} & 2.10\% \\
        & nDCG@10 & 0.1902 & 0.1334 & 0.2527 & 0.2390 & \underline{0.2869} & 0.2775 & \textbf{0.2991} & 4.25\% \\
        & MRR & 0.1660 & 0.1202 & 0.2191 & 0.2100 & \underline{0.2520} & 0.2378 & \textbf{0.2675} & 6.15\% \\
        \midrule
        \multirow{6}*{Toys} 
        & HR@1 & 0.0585 & 0.0257 & 0.0997 & 0.1114 & \underline{0.1878} & 0.1262 & \textbf{0.2075} & 10.49\% \\
        & HR@5 & 0.1977 & 0.0978 & 0.2795 & 0.2614 & \underline{0.3682} & 0.3344 & \textbf{0.3908} & 6.14\% \\
        & nDCG@5 & 0.1286 & 0.0614 & 0.1919 & 0.1885 & \underline{0.2820} & 0.2327 & \textbf{0.3115} & 10.46\% \\
        & HR@10 & 0.3008 & 0.1715 & 0.3896 & 0.3540 & \underline{0.4663} & 0.4493 & \textbf{0.4850} & 4.01\% \\
        & nDCG@10 & 0.1618 & 0.0850 & 0.2274 & 0.2183 & \underline{0.3136} & 0.2698 & \textbf{0.3353} & 6.92\% \\
        & MRR & 0.1430 & 0.0819 & 0.1973 & 0.1967 & \underline{0.2842} & 0.2338 & \textbf{0.3040} & 6.97\% \\
        \midrule
        \multirow{6}*{Yelp} 
        & HR@1 & 0.0801 & 0.0624 & 0.2053 & 0.2188 & 0.2375 & \underline{0.2405} & \textbf{0.2725} & 13.31\% \\
        & HR@5 & 0.2415 & 0.2036 & 0.5437 & 0.5111 & 0.5745 & \underline{0.5976} & \textbf{0.6202} & 3.78\% \\
        & nDCG@5 & 0.1622 & 0.1333 & 0.3784 & 0.3696 & 0.4113 & \underline{0.4252} & \textbf{0.4532} & 6.58\% \\
        & HR@10 & 0.3609 & 0.3153 & 0.7265 & 0.6661 & 0.7373 & \underline{0.7597} & \textbf{0.7755} & 2.08\% \\
        & nDCG@10 & 0.2007 & 0.1692 & 0.4375 & 0.4198 & 0.4642 & \underline{0.4778} & \textbf{0.5037} & 5.42\% \\
        & MRR & 0.1740 & 0.1470 & 0.3630 & 0.3595 & 0.3927 & \underline{0.4026} & \textbf{0.4306} & 6.95\% \\
        \bottomrule
    \end{tabular}
    \label{tab:results2}
\end{table*}

\section{Pipeline of \name Solution}\label{sec:A}

The whole pipeline of \name solution is described in Algorithm~\ref{alg:solution}.
Following such a process, one can easily train and deploy a \name model in a lightweight manner.

\section{Training Configuration}\label{sec:B}

LLMs are sensitive to the training configuration.
In order to get desirable results, we implement different configurations on different datasets as shown in Table~\ref{tab:config}.
Typically, different datasets require different training epochs and learning rates to achieve better performance (usually 2-5 epochs is enough) while the number of warmup steps is also important.
The batch size is better to be 16 while larger or smaller batch size may lead to failure in convergence.
\name is not so sensitive to the rank and alpha of LoRA but changing the dropout or modules of LoRA will lead to significant performance fluctuation.
We choose to use the cosine scheduler with a weight decay of 0.1 to control the learning rate in training process.

\section{Additional Results}\label{sec:C}

In addition to the main results, we also perform an evaluation with negative sampled items following the strategy of previous works~\cite{DBLP:conf/icdm/KangM18,DBLP:conf/cikm/ZhouWZZWZWW20,DBLP:conf/www/ZhouYZW22}.
Specifically, we adopt HR@1 (nDCG@1=HR@1), HR@5, HR@10, nDCG@5, nDCG@10 and MRR as evaluation metrics considering the much shorter candidate lists.

As illustrated in Table~\ref{tab:results2}, our proposed \name can still exceed all the baseline models with a significant margin.
Such results prove the superiority of \name on the ranking ability.